\documentstyle[pre,aps,multicol]{revtex}   
\input{psfig}

\begin{document}
\draft

\title{Anomalous roughening of wood fractured surfaces}

\author{St\'{e}phane Morel$,^1$ Jean Schmittbuhl$,^2$ Juan M. L\'{o}pez$,^3$ 
        and G\'{e}rard Valentin${\,}^1$}
\address{$^1$ Lab. de Rh\'{e}ologie du Bois de Bordeaux, 
        UMR 123, Domaine de l'Hermitage,
        B.P.10, 33610 Cestas Gazinet, France\\
        $^2$ Lab. de G\'{e}ologie, URA 1316, Ecole Normale Sup\'{e}rieure,
        24 rue Lhomond, 75231 Paris Cedex 05, France\\
        $^3$ Department of Mathematics, Imperial College, 180 Queen's Gate,
        London SW7 2BZ, United Kingdom}
        
\maketitle

\begin{abstract}
  Scaling properties of wood fractured surfaces are obtained from samples
  of three different sizes. Two different woods are studied: Norway spruce and
  Maritime pine. Fracture surfaces are shown to display an anomalous
  dynamic scaling of the crack roughness. This anomalous scaling behavior
  involves the existence of two different and independent roughness
  exponents.  We determine the local roughness exponents ${\zeta}_{loc}$ to
  be 0.87 for spruce and 0.88 for pine. These results are consistent with
  the conjecture of a universal local roughness exponent. The global
  roughness exponent is different for both woods, $\zeta$ = 1.60 for spruce
  and $\zeta$ = 1.35 for pine. We argue that the global roughness exponent
  $\zeta$ is a good index for material characterization.  
\end{abstract}
\begin{multicols}{2}      
\narrowtext              
\protect
\section{Introduction}
\label{sec:intro}
Since the pioneering work of Mandelbrot \textit{et. al.} \cite{Mandel}, it has
been firmly established that topography of fracture surfaces exhibits
remarkable scaling properties.
A fracture surface $z(x,y)$ is statistically invariant under an
anisotropic scaling transformation:
\begin{equation}
(x,y,z) \rightarrow (\lambda x,\lambda y,\lambda^{\zeta}z) \label{affine}
\end{equation}
where $\zeta$ is the \textit{roughness exponent}.  Experimental results
obtained on various materials (steels \cite{Mandel}, glass \cite{Dag1},
rocks \cite{Schmit1,Schmit2}, ceramics \cite{Mech,Mal}, metallic alloys
\cite{Mal,Bouch1,Plan} and aluminium alloys \cite{Dag2,Bouch2}), both
fragile and ductile, have shown that the roughness exponent $\zeta$ is
found between 0.7 and 0.9 (see \cite{Bouch3} for a recent review).  The
robustness of the results seems to support the idea suggested by Bouchaud
\textit{et. al.} \cite{Bouch2} that $\zeta \simeq 0.8$ might be a universal
value of the roughness exponent, \textit{i.e.}, independent of the material
properties.  This conjecture implies that the fracture toughness is not
correlated to the roughness exponent $\zeta$.  However, morphology of
fracture surfaces seems to be affected by material properties.

It has been suggested by Bouchaud \textit{et. al.} \cite{Boujp} that models
of front lines propagating through randomly distributed impurities
\cite{Halp,Ert1,Ert2} might be relevant to understand the morphology of the
fracture surfaces\cite{Schmit4,Rama}.  The development of the fracture
roughness has been described as a Family-Vicsek scaling \cite{Fam,Schmit1}.
However, in a very recent experimental study \cite{Lopez1}, it has been
found that the surface of a brittle fracture in a granite block exhibited
anomalous dynamic scaling properties akin to what occurs in some models of
nonequilibrium kinetic roughening \cite{Krug,Schro,Das,Lopez2,Lopez3}.

The anomalous scaling is defined as follows. 
The development of the
fluctuations of the height $h(x,t)$ with time is characterized by the root
mean square $w(l,t)$ at time $t$ over a window size $l$ along the $x$-axis
(perpendicular to the propagation direction)
\begin{equation}
w(l,t)={\biggl \langle \frac{1}{l} \sum_{i=1}^{l} h(x_i,t)^2 - \Bigl (\frac{1}{l} 
        \sum_{i=1}^{l} h(x_i,t)\Bigr )^2 \biggr \rangle}_{j}^{1/2} \label{rms}
\end{equation}
where the brackets $\langle ...\rangle_j $ denotes an average over the window
position $j$.  The roughness $w(l,t)$ is expected to scale in the case of
anomalous scaling as \cite{Lopez3} 
\begin{eqnarray}
w(l,t)\sim
\left\{ \begin{array}{ll}
        t^{{\beta}_{\ast}} \: l^{{\zeta}_{loc}} & \mbox{if \quad $l \ll t^{1/z}$} \\
        t^{\zeta /z} & \mbox{if \quad $l \gg t^{1/z}$}\label{local}
        \end{array}
\right.
\end{eqnarray}
where the exponent ${\beta}_{\ast}=(\zeta -{\zeta}_{loc})/z$ is an
anomalous time exponent.  This anomalous dynamic scaling involves two
different and independent roughness exponents: the local roughness exponent
${\zeta}_{loc}$, which describes the scaling when one considers windows
smaller than the system size, and the global exponent $\zeta$ for scaling
involving the system size\cite{Lopez3}.  The local roughness exponent
${\zeta}_{loc}$ is actually at reach of the methods currently used for
experiment analyses.  The global exponent $\zeta$ is more difficult to
extract from a classical roughness measurement.  Both exponents have to be
taken into account for a complete description of the scaling behavior of
the surface.  According to Eq.  (\ref{local}), the correlation 
length $\xi(t) \sim t^{1/z}$ corresponds to a characteristic lenght
below which the surface appears as self-affine with the
local exponent ${\zeta}_{loc}$.

In Ref.\cite{Lopez1}, global and local roughness exponents, 
$\zeta = 1.2$ and ${\zeta}_{loc} = 0.79$ respectively, 
have been measured . The latter
study was performed on a mechanically isotropic material (granite).
However, many materials have anisotropic mechanical properties like wood,
reinforced concrete and most composite materials.  Anisotropic properties
result generally from structural reinforcements along specific directions.
It is of great interest to understand how fractures in such materials  are
influenced by the anisotropic texture.  

In this study, we determine the complete scaling behavior of the fracture
roughness resulting from stable crack propagation in wood samples of
different sizes.  For two different woods (Maritime pine and Norway
spruce), we show that the local fluctuations of crack surfaces exhibit
anomalous dynamic scaling properties.  The global roughness exponent is
different for both woods. Local
roughness exponents are identical for both woods and support the conjecture
of a universal local roughness exponent for brittle fracture surfaces. The
main consequence of this anomalous scaling is that the magnitude of the
surface fluctuations over regions is not just a function of the region size
but also of the system size.

The paper is organized as follows. In Sec. \ref{sec:exp} we describe
experimental setups for crack propagation and fracture surface measurement.
Section \ref{sec:anomal} is devoted to the anomalous dynamic scaling
behavior.  In Sec. \ref{sec:consq}, we study the roughness magnitude as
function of the system size.  Finally, we discuss implications for fracture
process in Sec. \ref{sec:conclu}.
\protect
\section{Experiment}
\label{sec:exp}
Wood is a natural material which displays a structural anisotropy resulting
from the presence of running cells in the radial direction.  Two
commercially wood species have been tested : Maritime pine (\textit{Pinus
pinaster Ait}) and Norway spruce (\textit{Picea abies L.}).  Pine
specimens have an average oven dry specific weight $(\rho )$ of 560
kg/m$^3$ and growth rings are approximately 4 mm wide.  Typical values for
spruce specimens are: $(\rho )$ = 390 kg/m$^3$ and growth rings of 2 to 5
mm wide.  Moisture content of all specimens was measured between 11 and 13
\%.

Crack surfaces are obtained from a modified Tapered Double Cantilever Beam
(TDCB) specimens.  A fracture was initiated from a straight notch machined
with a band saw (thickness 2 mm) and prolonged on few millimeters with a
razor blade (thickness 0.2 mm).  Fracture is obtained through uniaxial
tension with a constant opening rate (Fig.\ref{fig1}).  The tapered shape
of the specimens allows to obtain a mode I stable crack growth (see Ref.
\cite{Liebo} for details) which induces a constant crack speed.  The crack
speed was around 0.6 mm/s (from 0.3 mm/s for small specimens to 1mm/s for
large specimens).  Crack surfaces were generated along an average
radial-longitudinal plane by aligning the growth rings perpendicular to the
straight notch.  In order to obtain an evolution of the amplitude of the
roughness as function of the system size, three geometrically similar
specimens of sufficiently different sizes have been fractured.  We used
samples of size $L$ equal to 11.25, 30, and 60 mm (see Fig.\ref{fig1}).

Anatomical characteristics of wood introduce typical scales which might
appears as cutoffs for scale invariances.  Most tetragonal tracheid cells
in pine and spruce are about 25 $\mu$m wide.  During loading cell walls
break revealing U-shaped profiles with rugged edges because of the
rectangular shape of the tracheid section.  Thickness of cell walls varies
from 2 to 10 $\mu$m.  This facies of fracture surface is characteristic of
a local brittle fracture process.

Topographies of the crack surfaces were recorded with a mechanical profiler
along regular grids. Grid axes are along the $x$ direction which is
parallel to the initial notch and along the $y$ direction which is the
crack propagation direction (Fig.\ref{fig1}).  The step of sampling in the
$x$ direction is adjusted to the minimum cell width : $\Delta x$ = 25
$\mu$m and to the cell length in the $y$ direction : $\Delta y$ = 2.5 mm.
Profiles along the $x$-axis were sampled with 2050 points for specimens of
width $L$ = 60 mm, 1030 points for specimens of width $L$ = 30 mm, and 360
points for specimens of width $L$ = 11.25 mm.  For each map, the first
profile ($y$ = 0) is sampled in the immediate vicinity of the initial
straight notch and has a zero roughness.  As the distance $y$ to the notch 
increases, the roughness develops up to 3 mm.  The vertical resolution is
estimated from the height differences between two successive sampling along
the same line. Its magnitude is about 3 $\mu$m.  Horizontal resolutions
along $x$ and $y$-axis are about 5 $\mu$m.  In the case of pine an
additional specimen size was tested : $L$ = 22.50 mm with 800 points, but
only profiles far from the notch have been recorded.  Table \ref{tab1}
lists parameters of the various studied samples.
\protect
\section{Anomalous dynamic scaling}
\label{sec:anomal}
As mentionned above, fractures of all specimens have been obtained at
constant crack speed.  Subsequently, we assumed a linear relationship
between the $y$-position of the profiles and the crack propagation time
$t$.  Height profiles are considered as descriptions of the advancing crack
front $h(x,t)$. Complete spatio-temporal evolution of the crack front can
thus be produced from roughness maps.

In Figure \ref{fig2}, we present the development of the roughness $w(l,t)$
versus time $t$ in a log-log plot for different window sizes $l$ in the
case of the spruce specimen (s60-1) which is 60 mm wide.  The upper line is
a fit of the roughness growth for the largest window size ($l = 13.975$
mm). The slope of this fit provides an estimate of the ratio of the global
roughness exponent and the dynamical exponent : $\zeta /z \approx 0.26$.
The fit is computed for times between time $t_{min}$ and time $t_{max}$.
Before time $t_{min}$ the crack speed is not constant. After time
$t_{max}$, the roughness has saturated because of the reach of the system
size.

The lower line is a fit of the roughness measured for a small window size
($l$ = 0.175 mm).  It appears that $w(l,t)$ increases like a power law as a
function of the crack propagation time $t$ even for small window sizes. The
slope of the fit is 0.14 significantly larger than zero and gives an
estimate of the $\beta_*$ exponent.  This unconventional dependence on time
is an illustration of the anomalous scaling and differ from a Family-Vicsek
scaling where the roughness is expected to be time independent for small
window sizes.

These two regimes are in good agrement with the anomalous scaling proposed
in Eq.(\ref{local}).  A similar behavior has been observed for all
specimens.
\protect
\subsection{Local roughness exponent}
\label{sec:loc-exp}
The local roughness exponent ${\zeta}_{loc}$ is determined using four
methods: the variable bandwidth methods : root mean square and max-min
difference\cite{Schmit1,Schmit3}, the power spectrum and the averaged
wavelet coefficient method\cite{Simon}.  Local roughness exponents
${\zeta}_{loc}$ have been determined on profiles located far from the
notch, \textit{i.e.}, at long times. Results on specimen s60-1 are used as
illustrations. Complete results for all specimens are provided in
Table~\ref{tab1}.
  
In the root mean square method, the roughness $w$ over a window $l$ 
is expected from Eq.(\ref{rms}) to scale at long enough time as
\begin{equation}
  w(l,t\gg l^{1/z}) \sim l^{\zeta_{loc}}
\end{equation}
From Figure \ref{fig3} the local roughness exponent is :  ${\zeta}_{loc}$ =
0.84 in the case of specimen s60-1.

The max-min method consists of the computation of $h_{max}(r)$, which is
defined as the difference between the maximum and the minimum heights $h$
within this window, averaged over all possible origins $x$ of the window
\cite{Fed}: $ h_{max}(r) = {\langle Max\{h(r')\}_{x<r'<x+r} -
  Min\{h(r')\}_{x<r'<x+r} \rangle}_{x}$. For a self-affine profile,
$h_{max}$ is expected to scale as:
\begin{equation}
h_{max}(r) \sim  r^{{\zeta}_{loc}} \label{max}
\end{equation}
where $r$ is the width of the window along $x$-axis.  For specimen s60-1,
we measured a local roughness exponent:  ${\zeta}_{loc}$ = 0.89.

The third method is a calculation of the power spectrum, \textit{i.e.}, the
Fourier transform of the autocorrelation function $\langle h(x+\Delta
x)h(x)\rangle$.  The power spectrum scales for a self-affine profile as\cite{Fed}:
\begin{equation}
S(k) \sim  k^{-(2{\,}{\zeta}_{loc}+1)} \label{fft}
\end{equation}
where $k$ is the wave factor.  In Figure \ref{fig4} we show a log-log
plot of $S(k)$ versus $k$ for specimen s60-1.  $S(k)$ decays with a power
law $k^{-2.78}$ which is consistent with ${\zeta}_{loc}$ = 0.89.

The last method used in this study is the averaged wavelet coefficient
method\cite{Simon}.  This method consists of the average of the wavelet transform of
the profile over the translation factor $b$.  The averaged wavelet
coefficient $W[h](a)$ scales as
\begin{equation}
W[h](a) \sim  a^{\frac{1}{2} +{\zeta}_{loc}} \label{wav}
\end{equation}
where a is the scale factor.

The estimates of the local roughness exponents obtained with these four
methods for all specimens are given in Table \ref{tab1}.  As shown in Table
\ref{tab1}, the values of the local roughness exponent ${\zeta}_{loc}$
calculated by the root mean square method, the max-min method and the power
spectrum decrease with the system size $L$. Only values obtained from the
wavelet analysis seem independent of the system size.  In the following, we
show that this deviation is due to measurement and analysis biases and can
be corrected.

The reliability of the determination of self-affine exponents has already been
studied \cite{Simon,Schmit3}.  It has been shown that several
artifacts may introduce systematic errors for the estimation of the
local roughness exponent. Two types of biases have to be distinghished: those
which happen during the geometric measurement of the object and those which
are relative to the method of signal analysis.

In our study, profiles are recorded with a needle moving along crack
surfaces. For a similar type of measurement \cite{Schmit2,Schmit3}, it has
been shown that shape and volume of the needle can induce a geometric
filter.  When the tip of the needle is a half-sphere, the needle follows
hills more correctly than sharp holes.  Subsequently the exactness of the
measured height is function of the surroundings.  It has been found that an
increase of the radius of the needle tip induces an increase of the
measured roughness exponent (see \cite{Schmit3} for more details).

In the case of biases relative to the analysis methods, it has been found
that the accuracy of the different methods is sensitive to two parameters :
the size of the system (number of recorded points) and the roughness
exponent.  In our study, the system size strongly evolves from small to big
specimens : 360 to 2050 points.  In Ref.~\cite{Schmit3}, tests on synthetic
profiles generated with self-affine exponent between 0.8-0.9 show that the
three methods underestimate the self-affine exponent. The underestimation
is larger when the system size decreases. The root mean square method is
the most sensitive to this size effect.

Likely both biases exist in this study.  In order to evaluate simultanously
the influence of both flaws on local roughness exponents, synthetic
profiles are simulated, filtered and analysed.  Self-affine profiles are
simulated numerically with a \textit{Voss construction}~\cite{Voss} for
four values of self-affine exponent : 0.80, 0.85, 0.90 and 0.95.  For each
exponent, 100 independent profiles are generated. The horizontal step
between two consecutive points is $So$ = 6.25 $\mu$m corresponding to the
lower cutoff, \textit{i.e.}, the mean thickness of cell walls.
Magnification of self-affine profiles corresponds to that measured on
experimental profiles.  The filter is an under sampling with a sphere of
radius $R$ = 25 $\mu$m (\textit{i.e.} size of the experimental needle)
every four steps ($S = 4So = 25 \mu$m).  The step $S$ corresponds to the
experimental step $\Delta x$.  Output exponents are obtained with the four
methods (rms, max-min, power spectrum and wavelet analysis) and are given
in Table \ref{tab2} for different system sizes.  From Table \ref{tab2}, the
corrected values of the experimental ${\zeta}_{loc}$ are estimated and
given in brackets in Table \ref{tab1}.  Average of the corrected values of
${\zeta}_{loc}$ obtained from the different methods gives the local
roughness exponents : $0.87{\pm}0.07$ for spruce and $0.88{\pm}0.07$ for
pine.  Results are consistent with those obtained for brittle materials
where ${\zeta}_{loc} \approx 0.85$ \cite{Mal,Schmit2,Bouch3} and support
the conjecture of a universal local roughness exponent.

Our results are different from those obtained by Eng\o y \textit{et al}
\cite{Eng} who studied the roughness of brittle fractures for different
woods. The authors found a local roughness exponent ${\zeta}_{loc}$ = 0.68
which is characteristic of a two-dimensional fracture. Several reasons
might explain this difference. First, direction of propagation crack was
perpendicular to fibers while in our setup propagation is parallel to
fibers.  Second moisture content of tested specimens was around 4\% which
is significantly lower than that measured in our study (12\%).  Low
moisture content induces micro-cracking in the radial-longitudinal and
tangential-longitudinal planes of wood due to drying shrinkage.  This
mechanism of micro-cracking does not appear in mode I fracture.
Micro-cracks induce preferential paths for the macro-crack which modify the
scaling properties of fracture surfaces.  Third the experimental procedure
was strongly different in the study of Eng\o y \textit{et al} since
fracture propagation was unstable contrary to the stable propagation in the
present work.
\protect
\subsection{Global roughness and dynamical exponents}
\label{sec:glob-dyn}
As discussed above, the existence of an exponent $\beta_{\ast} \ne 0$ (see
Fig. 2) indicates that an anomalous roughening is taking place.  To obtain
an accurate description of the anomalous scaling, we follow
\cite{Lopez1,Lopez3} and define the scaling function $g(u)$ as
$g({\,}l/{\,}t^{1/z}) = w(l,t)/{\,}l^{\zeta} $.  From Eq.(\ref{local}),
$g(u)$ is expected to scale like
\begin{eqnarray}
g(u)\sim
\left\{ \begin{array}{ll}
        u^{-({\zeta}-{\zeta}_{loc})} & \mbox{if \quad $u \ll 1$} \\
        u^{-\zeta} & \mbox{if \quad $u \gg 1$}\label{scal}
        \end{array}
\right.
\end{eqnarray}
The scaling function $g$ is computed by data collapses from each profile of
a complete crack map, (\textit{i.e.} the set of profiles that describe a
single fracture). In Figures \ref{fig5} and \ref{fig6} we present the data
collapses of $g(u)$ for all the maps obtained for the three specimen sizes
($L$ = 60, 30 and 11.25 mm) of both spruce and pine.

Fig.\ref{fig5}(a) is considered as a good example of these data collapses.
The quality of the collapse is used for the determination of the dynamical
exponent $z$. The global exponent $\zeta$ is obtained from the fit of the
scaling function.  The time evolution of the height fluctuations at small
scales is shown by the nonconstant behavior for $u\ll1$.  This regime is
fitted by a power law $g(u)\propto u^{-0.76}$.  Using our previous estimate
of the local roughness exponent ${\zeta}_{loc} = 0.84$ we obtain the
magnitude of the global roughness exponent $\zeta = 1.60$.  In the
particular case of sample s60-1 shown in Fig. \ref{fig5}(a) the best
collapse is observed for the dynamical exponent $z = 5.9$.  Note that
estimates of the exponents ${\zeta}_{loc} = 0.84$, $\zeta = 1.60$ and $z =
5.90$ are very consistent with fits obtained from Fig.~\ref{fig2}:
${\beta}_{\ast} = 0.13$ and ${\zeta}/z = 0.27$ for this sample.

Data collapses of all maps are presented in Fig.\ref{fig5} and 
Fig.\ref{fig6} and
are in good agreement with a scaling function like (\ref{scal}).  For both
wood species, global roughness exponent and dynamical exponent are reported
in Tab.\ref{tab1}.  As shown in Tab.\ref{tab1}, global roughness exponents
${\zeta}$ are independent of the system size.  Average values are 
${\zeta} = 1.60\pm 0.10$ for spruce and ${\zeta} = 1.35\pm 0.10$ for pine.
\protect
\section{Implications of anomalous scaling}
\label{sec:consq}
According to Eq.(\ref{local}), the roughness is expected to 
saturate only at
times $t \gg L^{z}$, \textit{i.e.}, when the correlation length $\xi (t)
\sim t^{1/z}$ has reached the boundary length,  ${\xi}_{max} \propto L$.
In this regime the roughness magnitude scales with the system size for any
window length even much smaller than the system size $L$:
\begin{equation}
w(l,t \gg L^{z}) \sim  l^{{\zeta}_{loc}}\: L^{{\zeta}-{\zeta}_{loc}} \label{size}
\end{equation}
We checked the linear relationship between ${\xi}_{max}$ and the system size
$L$ by measuring ${\xi}_{max}$. From the evolution of the roughness
$w(l,t)$ with time (see Fig.~\ref{fig2}), the saturation time $t_{sat}$ is
estimated. The correlation length $\xi_{max}$ is obtained using the
relation ${\xi}_{max} \propto t_{sat}^{1/z}$. Values of ${\xi}_{max}$ for the
different maps are reported in Table~\ref{tab1}.  In figure \ref{fig7},
${\xi}_{max}$ is plotted versus $L$ for both woods.  A linear relationship
between ${\xi}_{max}$ and the system size $L$ exists except in the case of
spruce for the biggest sample size where the saturation regime is not
clearly reached.

In Figure \ref{fig8}, the ratio ${ \langle w(l,t \gg \xi_{max}^z){\,}
  l^{{\zeta}_{loc}} \rangle}_{l}$ is plotted versus $L$ for profiles at
times $t \geq ({\xi}_{max})^{z}$ for pine specimens.  A power law
$L^{{\zeta}-{\zeta}_{loc}}$ with exponents determined previously $\zeta =
1.35$ and ${\zeta}_{loc} = 0.80$ is very consistent with data. It confirms
the increase of the roughness magnitude with the system size $L$ even for
windows smaller than the system size.  In Figure \ref{fig8} 
${ \langle w(l \ll {\xi}_{max},t)/({\,}l^{{\zeta}_{loc}}\:
  {{\xi}_{max}}^{{\zeta}-{\zeta}_{loc}}) \rangle}_{l}$ versus $L$ 
is also plotted (filled symbols) which is expected to be constant.
\protect
\section{Conclusions}
\label{sec:conclu}   
In this study we have shown that fracture surfaces of an anisotropic
material like wood display an anomalous dynamic scaling of the crack
roughness.  From different specimen sizes, we have studied the size effects
on roughness exponents.  It appears that the global roughness exponent is
independent of the system size and different for both studied woods.  We
have obtained ${\zeta} = 1.60\pm 0.10$ for spruce and ${\zeta} = 1.35\pm
0.10$ for pine.  The local roughness exponent ${\zeta}_{loc}$ shows a
deviation according to the system size.  However, we argue that this
deviation is due to a biased estimate resulting from two independent
effects: the number of sampled points and the local filtering resulting
from the needle shape during the roughness measurement.  Errors due to
these biases have to be considered and the corrected values of the local
roughness exponents are $0.87{\pm}0.07$ for spruce and $0.88{\pm}0.07$ for
pine.  These results support the conjecture of a universal local roughness
exponent for brittle materials.  Moreover, we have shown that it exists a
linear relationship between the system size and the maximum 
correlation length ${\xi}_{max}$. This relation induces a
system size dependence in the roughness magnitude at saturation.

Our results can be compared with a recent experiment in granite 
\cite{Lopez1} in which the exponents $\zeta_{loc} = 0.79$ and
$\zeta = 1.2$ were obtained. We suggest that 
the global roughness exponent, which 
seems to be dependent of material, may be a good candidate 
as an index for characterizing material properties. On the contrary, the
local roughness exponent does not seem to change for different
materials and might be {\em universal}. 
Up to our knowledge, the existing models of cracks 
are based on the assumption that cracks are truly self-affine,{\it i.e.}
$\zeta = \zeta_{loc}$. It is a major interest
to find theoretical models of crack interfaces that could
incorporate anomalous kinetic roughening 
in a simple way. 

\acknowledgments
S.M. wishes to thank E. Bouchaud for very fruitful 
discussions and encouragement. 
J.M.L. also thanks M.A. Rodr\'{\i}guez for a careful reading
of the manuscript and the European Commission for support.
\begin{figure}
\protect
\caption{Modified tapered double cantilever beam (TDCB) 
specimen subjected to mode I 
crack propagation. The crack plane is perpendicular 
to the tensile axis which 
corresponds to radial-longitudinal plane of 
wood (longitudinal direction being the 
direction of crack propagation). Dimensions are given in mm.}
\label{fig1}
\end{figure}
\begin{figure}
\protect
\caption{Roughness (\textit{rms}) $w(l,t)$ \textit{vs.} time for a 
  spruce specimen 
  60 mm wide (s60-1) calculated over windows $l$ 
  of size ranging from $l$ = 0.175 mm to $l$
  = 13.975 mm with size step $\Delta l$ = 0.100 mm.  The
  continuous line (a) corresponds to the fit of the roughness between
  $t_{min}$ and $t_{max}$, $w(l,t) \sim t^{{\beta}_{\ast}}$ for a small
  window size $l$ = 0.175 mm.  ${\beta}_{\ast}$ = 0.14 is obtained.  The
  continuous line (b) is the fit of data for large window size $l$ = 13.975
  mm.  Its slope 0.26 corresponds to ${\zeta /z}$.}
\label{fig2}
\end{figure}
\begin{figure}
\protect
\caption{Roughness (\textit{rms}) $w(l)$ \textit{vs.} $l$ for the profile
  at the saturated time $t\gg{{\xi}_{max}}^z$ (specimen s60-1).  The straight
  line corresponds to the power law $w(l)\: \sim \: l^{{\zeta}_{loc}}$ and
  gives a determination of the local roughness exponent ${\zeta}_{loc}$ =
  0.84.}
\label{fig3}
\end{figure}
\begin{figure}
\protect
\caption{Power spectrum at time $t_{sat}$ in the case of s60-1 specimen.
  The straight line has slope -2.78 which is consistent with a power law
  $k^{-(2{\,}{\zeta}_{loc}+1)}$ and a local roughness exponent
  ${\zeta}_{loc}$ = 0.89.}
\label{fig4}
\end{figure}
\begin{figure}
\protect
\caption{Data collapses for spruce specimens data in three different 
  system sizes. 
  Panel (a), (b), (c) display the data collapses of s60-1, s30-1 and s11-1
  specimens which have respectively a size $L$ = 60, 30 and 11.25 mm.  The
  non-constant behavior (\textit{i.e.} non zero slope) at small values of
  $l/t^{1/z}$ displays the dependence on time of the roughness magnitude,
  Scalings are in good agreement with the scaling function (Eq.(\ref{scal})).}
\label{fig5}
\end{figure}
\begin{figure}
\protect
\caption{Data collapses for pine specimens of three different system sizes. 
  Panel (a), (b), (c) display the data collapses of p60-1, p30-2 and p11-2
  specimens having respectively a size $L$ = 60, 30 and 11.25 mm.}
\label{fig6}
\end{figure}
\begin{figure}
\protect
\caption{Maximum self-affine correlation lengths 
  ${\xi}_{max}$ \textit{vs.} the system 
  size $L$ for spruce (circle) and pine (square) specimens.  Both spruce
  and pine show a linear relationship (dashed line) between
  ${\xi}_{max}$ and $L$. The determination of ${\xi}_{max}$ in the case of
  spruce and large system size is underestimated owing to the to little
  duration of the roughness map. The saturation transition was not clearly
  observable for this sample.}
\label{fig7}
\end{figure}
\begin{figure}
\protect
\caption{Size effect on the amplitude of the roughness over 
profiles at saturation, \textit{i.e.}, at times 
$t \geq {{\xi}_{max}}^z$, for pine specimens. Upper symbols
correspond to ${ \langle w(l \ll {\xi}_{max},t)/{\,} 
l^{{\zeta}_{loc}} \rangle}_{l}$ versus $L$ : Circle correspond to 
the specimen of $L$ = 60 mm, squares $L$ = 30 mm, 
diamonds $L$ = 22.5 mm and triangles $L$ = 11.25 mm. 
Filled symbols are obtained for 
${ \langle w(l \ll {\xi}_{max},t)/({\,}l^{{\zeta}_{loc}}\: 
{{\xi}_{max}}^{{\zeta}-{\zeta}_{loc}}) \rangle}_{l}$ versus $L$.}
\label{fig8}
\end{figure}
\end{multicols}  
\newpage  
\widetext
\begin{table}
%
\caption{Description of the analysed specimens. 
Local roughness exponents are calculated 
using the root mean square method, the max-min difference method, 
the power spectrum,  and the averaged wavelet coefficient analysis.  
Values in brackets are corrected 
from errors due to measurement and analysis biases.}
\label{tab1}
\begin{tabular}{|c|c c c|c c c c c c c|}
 & Specimen & $L$ & Nb. of  & root mean & Power & max-min & wavelet& $\zeta$
 & $z$ & ${\xi}_{max}$ \\
species & label & (mm)& profiles & square & spectrum & & analysis 
&  &  & (mm)\\
\hline 
        & s60-1 & 60 & 49 & 0.84 (0.95) & 0.89 (0.86) & 0.89 (0.90) 
& 1.00 (0.96) & 1.60 & 5.90 & 3.90 \\
        & s60-2 & 60 & 49 & 0.81 (0.88) & 0.85 (0.82) & 0.89 (0.90) 
& 0.91 (0.87) & 1.55 & 2.40 & 4.30 \\
        & s60-3 & 60 & 43 & 0.84 (0.95) & 0.95 (0.93) & 0.87 (0.87) 
& 0.99 (0.95) & 1.60 & 5.60 & not sat. \\
spruce  & s30-1 & 30 & 48 & 0.78 (0.84) & 0.83 (0.80) & 0.83 (0.79) 
& 0.92 (0.87) & 1.55 & 3.50 & 4.10 \\
        & s30-2 & 30 & 47 & 0.79 (0.85) & 0.84 (0.81) & 0.82 (0.79) 
& 0.89 (0.84) & 1.60 & 2.00 & 4.20 \\
        & s11-1 & 11.25 & 22 & 0.73 (0.80) & 0.84 (0.85) & 0.83 (0.84) 
& 0.91 (0.84) & 1.55 & 2.50 & 1.35 \\
        & s11-2 & 11.25 & 25 & 0.77 (0.85) & 0.84 (0.85) & 0.83 (0.84) 
& 0.98 (0.91) & 1.60 & 2.60 & 1.45 \\
        & s11-3 & 11.25 & 26 & 0.79 (0.88) & 0.88 (0.93) & 0.84 (0.85) 
& 0.95 (0.88) & 1.55 & 2.60 & 1.40 \\
\hline
spruce & \multicolumn{3}{r|}{} & 0.88 ${\pm}0.05$ 
& 0.86 ${\pm}0.06$ & 0.85 ${\pm}0.07$ & 0.89 ${\pm}0.09$ & 1.60 ${\pm}0.10$ &
 \\
\hline
        & p60-1 & 60 & 45 & 0.84 (0.95) & 0.91 (0.88) & 0.90 (0.93) 
& 0.97 (0.93) & 1.30 & 1.90 & 7.30 \\
        & p60-2 & 60 & 46 & 0.81 (0.88) & 0.86 (0.83) & 0.88 (0.88) 
& 0.90 (0.86) & 1.35 & 2.30 & not sat. \\
        & p30-1 & 30 & 21 & 0.84 (0.95) & 0.87 (0.85) & 0.88 (0.88) 
& 0.99 (0.95) & 1.35 & 2.20 & 3.85 \\
        & p30-2 & 30 & 30 & 0.80 (0.87) & 0.81 (0.79) & 0.87 (0.87) 
& 0.99 (0.95) & 1.30 & 4.30 & not sat. \\
pine    & p30-3 & 30 & 31 & 0.85 (0.95) & 0.83 (0.80) & 0.90 (0.95) 
& 0.97 (0.93) & 1.40 & 2.60 & 3.80 \\
        & p30-4 & 30 & 31 & 0.83 (0.92) & 0.88 (0.88) & 0.89 (0.90) 
& 1.01 (0.96) & 1.35 & 3.90 & not sat. \\
        & p11-1 & 11.25 & 26 & 0.75 (0.83) & 0.86 (0.88) & 0.83 (0.83) 
& 1.03 (0.96) & 1.35 & 3.20 & 1.40 \\
        & p11-2 & 11.25 & 27 & 0.75 (0.83) & 0.86 (0.89) & 0.83 (0.83) 
& 0.98 (0.91) & 1.40 & 1.80 & 1.80 \\
        & p11-3 & 11.25 & 28 & 0.75 (0.83) & 0.82 (0.79) & 0.84 (0.84) 
& 0.97 (0.90) & 1.30 & 2.30 & not sat. \\
\hline
pine    & p22-1 & 22.50 & & 0.81 (0.89) & 0.84 (0.83) & 0.85 (0.83) 
& 0.99 (0.94) & & & 3.00\\
        & p22-2 & 22.50 & & 0.81 (0.89) & 0.86 (0.87) & 0.85 (0.83) 
& 0.94 (0.88) & & & 2.70\\
\hline
pine & \multicolumn{3}{r|}{} & 0.89 ${\pm}0.05$ &
$ 0.84 {\pm}0.07$ & 0.87 ${\pm}0.05$ & 0.92 ${\pm}0.08$ & 1.35 ${\pm}0.10$ &
  \\     
\end{tabular}
\end{table}
\begin{table}
%
\caption{Tests of root mean square, power spectrum, max-min difference 
and wavelet analyses on undersampled and filtered synthetic self-affine profiles 
which model profiler recording (see text for details).  
Four system sizes, in terms of number of points are considered.  
The accuracy of the exponents presented in this table is
around 8\%.}
\label{tab2}
\begin{tabular}{c|c c c c|c c c c|c c c c|c c c c}
 System size & \multicolumn{4}{c|}{256 pts} & \multicolumn{4}{c|}{512 pts} & 
 \multicolumn{4}{c|}{1024 pts} & \multicolumn{4}{c}{2048 pts} \\
\hline 
 self-afine exponent & 0.80 & 0.85 & 0.90 & 0.95 & 0.80 & 0.85 & 0.90 & 0.95 &
 0.80 & 0.85 & 0.90 & 0.95 & 0.80 & 0.85 & 0.90 & 0.95 \\
\hline
\hline
\textit{rms} & 0.72 & 0.77 & 0.80 & 0.82 & 0.74 & 0.78 & 0.81 & 0.83 &
 0.75 & 0.79 & 0.82 & 0.84 & 0.75 & 0.79 & 0.82 & 0.84 \\
\hline
power spectrum & 0.82 & 0.84 & 0.87 & 0.89 & 0.82 & 0.85 & 0.88 & 0.90 &
 0.83 & 0.87 & 0.89 & 0.91 & 0.83 & 0.88 & 0.93 & 0.96 \\
\hline
\textit{max-min} & 0.81 & 0.84 & 0.86 & 0.89 & 0.83 & 0.85 & 0.88 & 0.90 &
 0.84 & 0.86 & 0.89 & 0.90 & 0.84 & 0.86 & 0.89 & 0.91 \\ 
\hline
wavelet analysis & 0.87 & 0.92 & 0.97 & 1.02 & 0.87 & 0.91 & 0.96 & 1.00 &
 0.85 & 0.90 & 0.94 & 0.99 & 0.84 & 0.89 & 0.94 & 0.99 \\ 
\end{tabular}
\end{table}

\psfig{figure=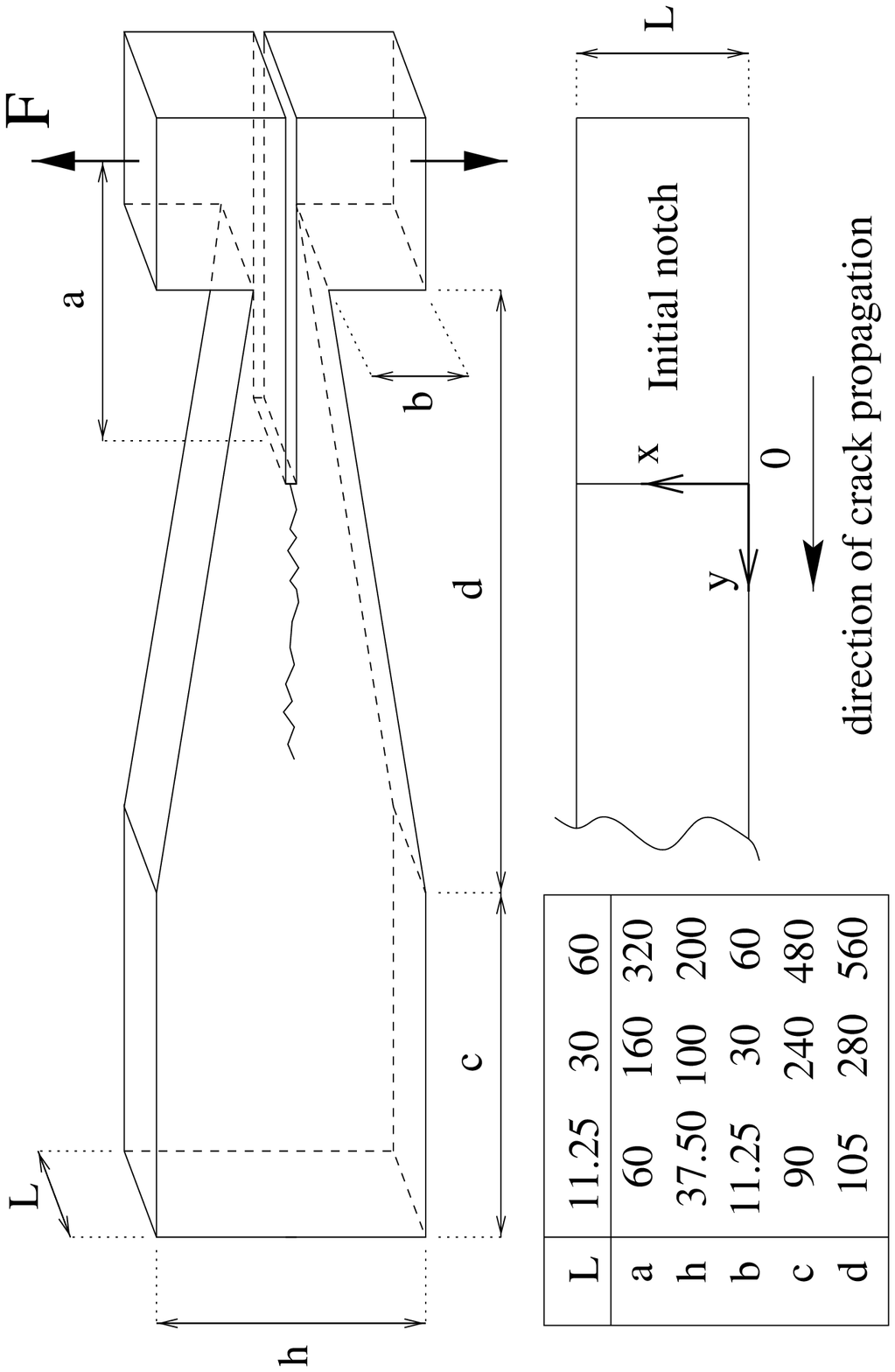}
\psfig{figure=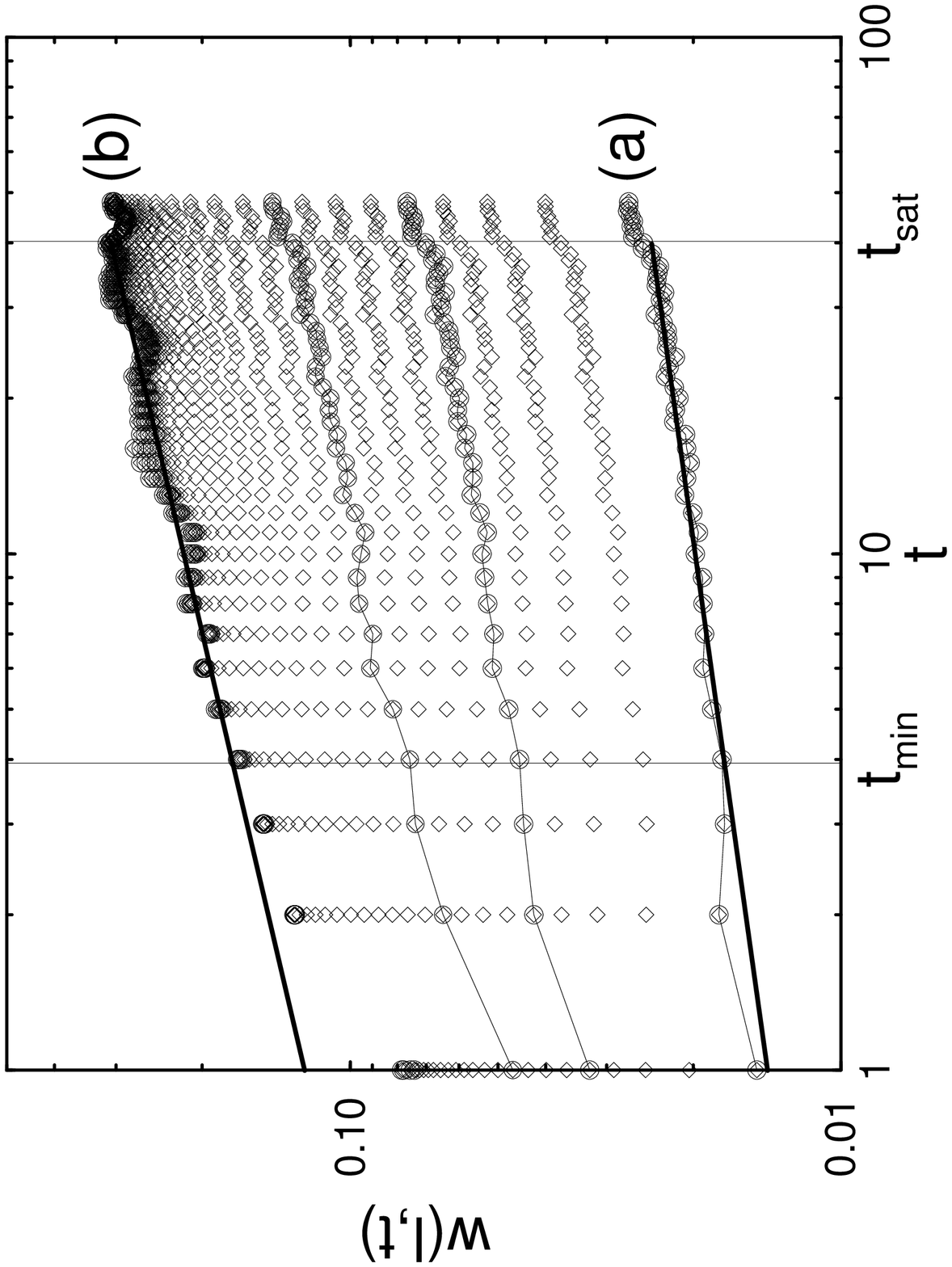}
\psfig{figure=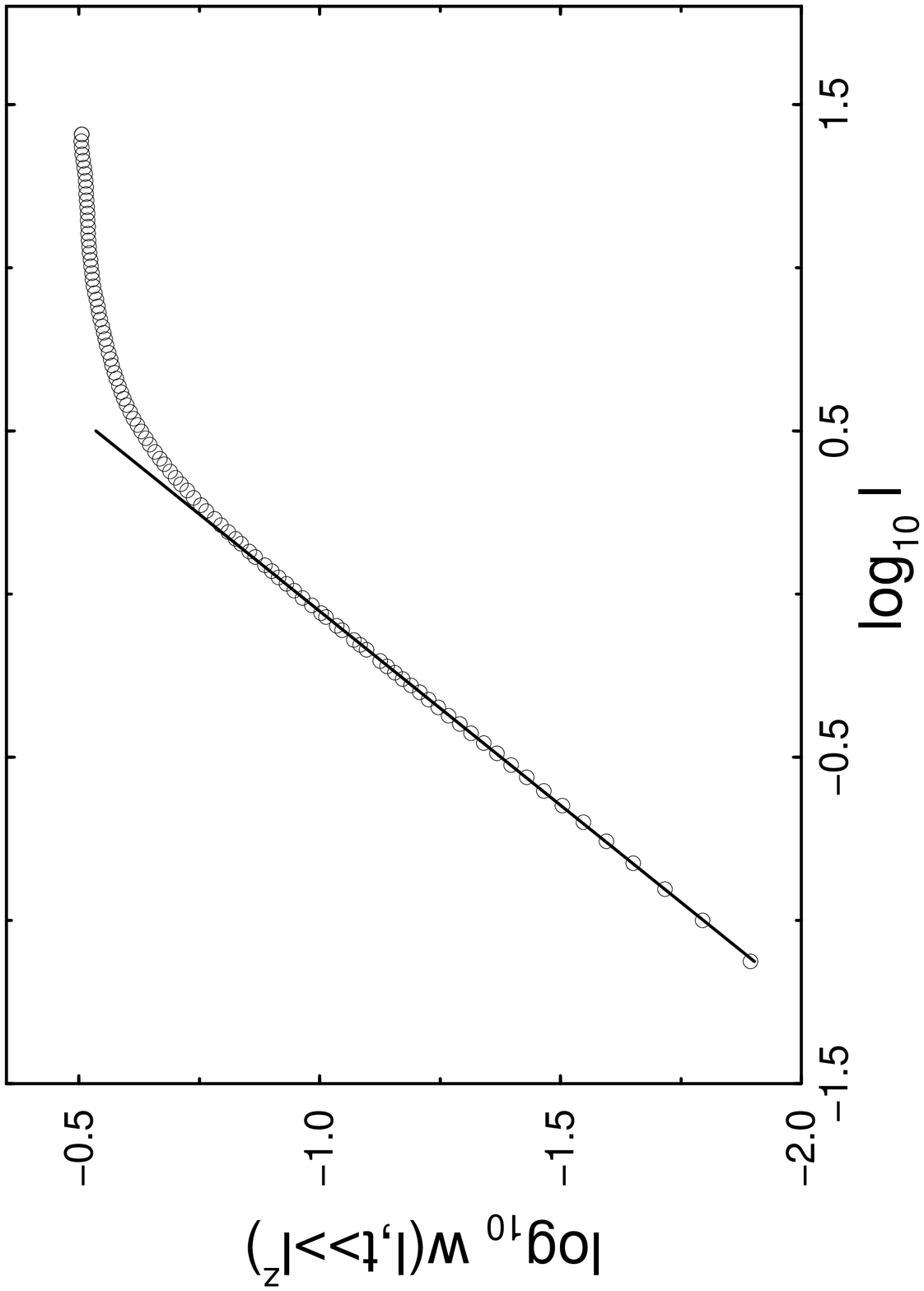}
\psfig{figure=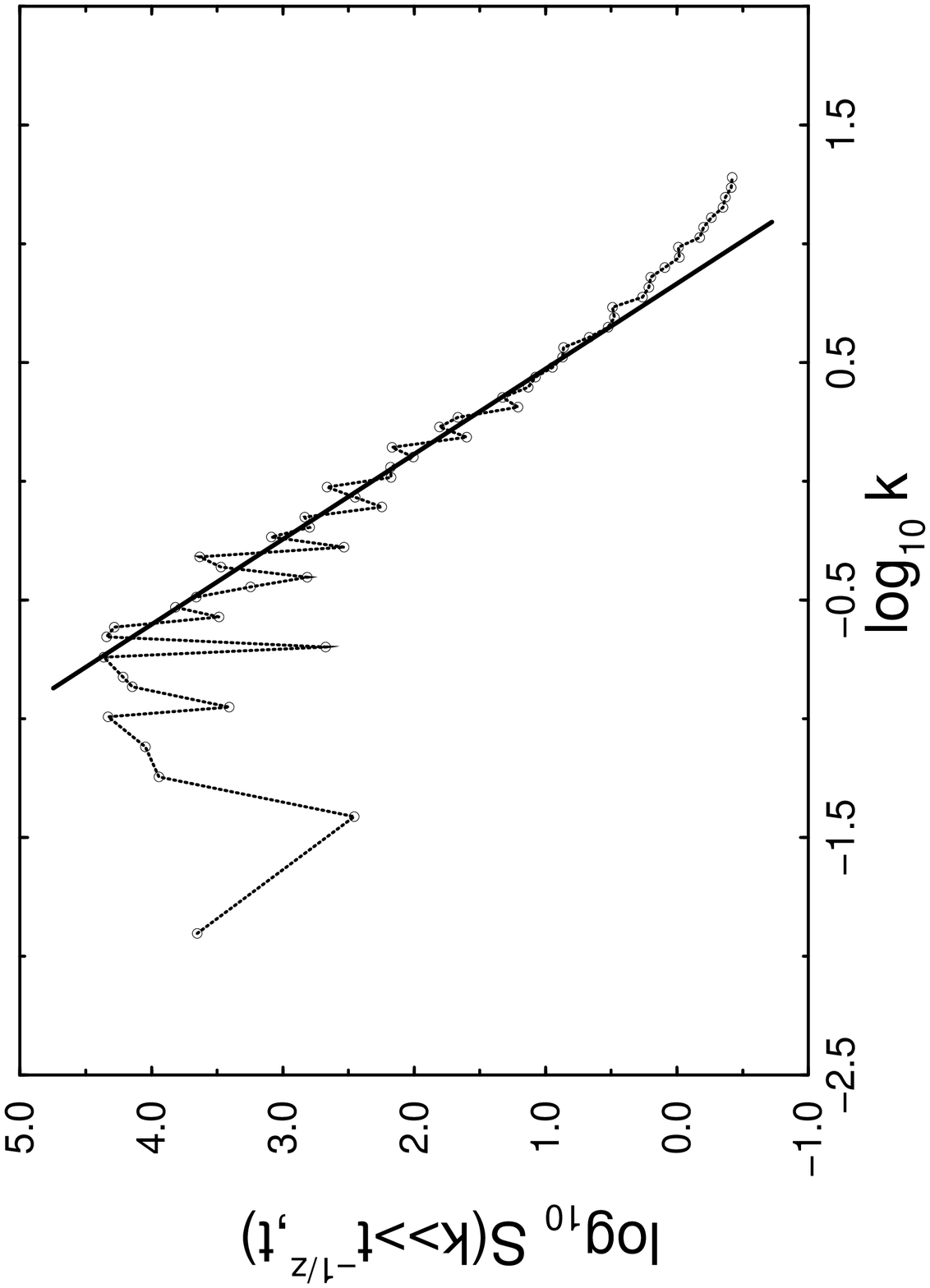}
\psfig{figure=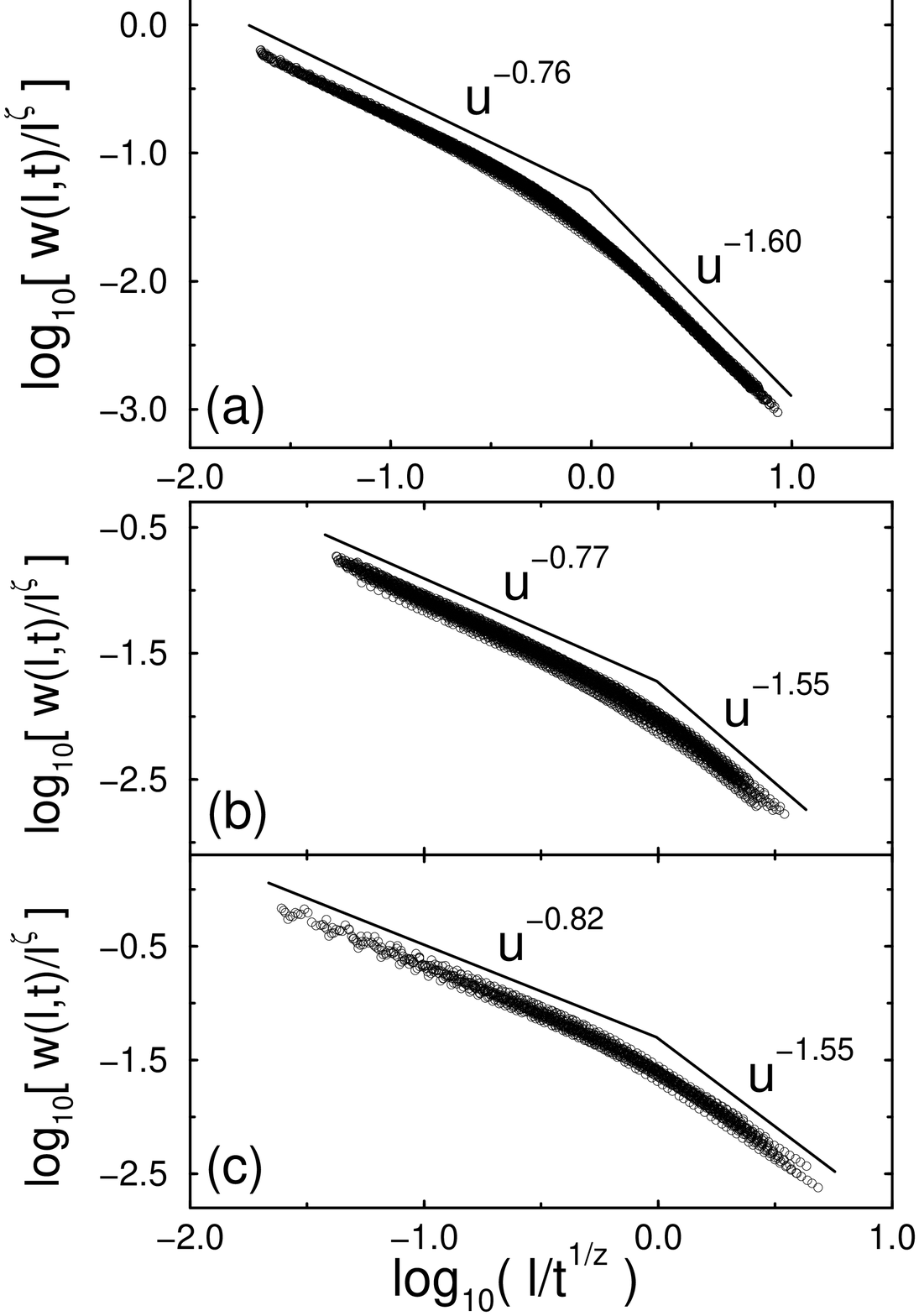}
\psfig{figure=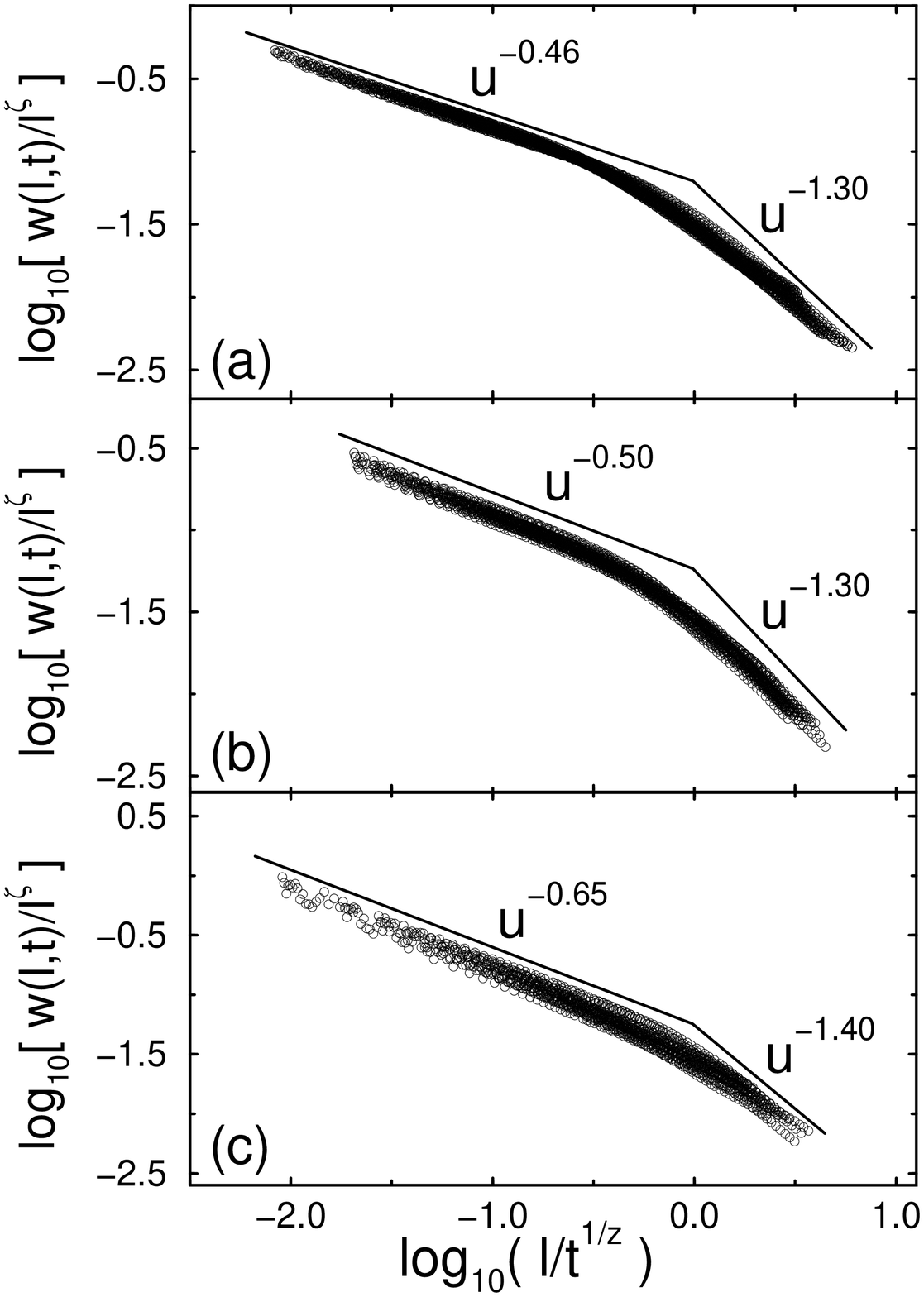}
\psfig{figure=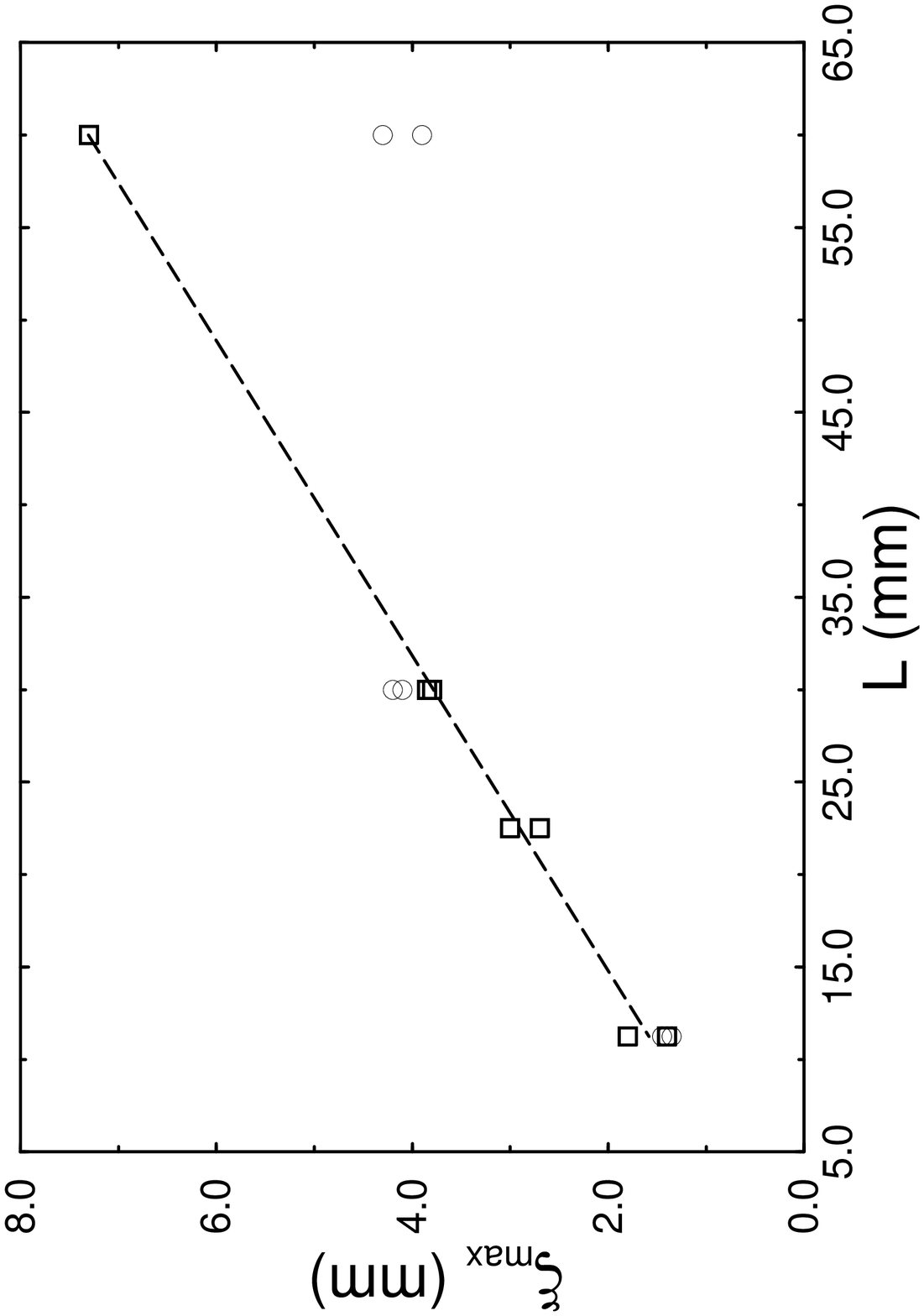}
\psfig{figure=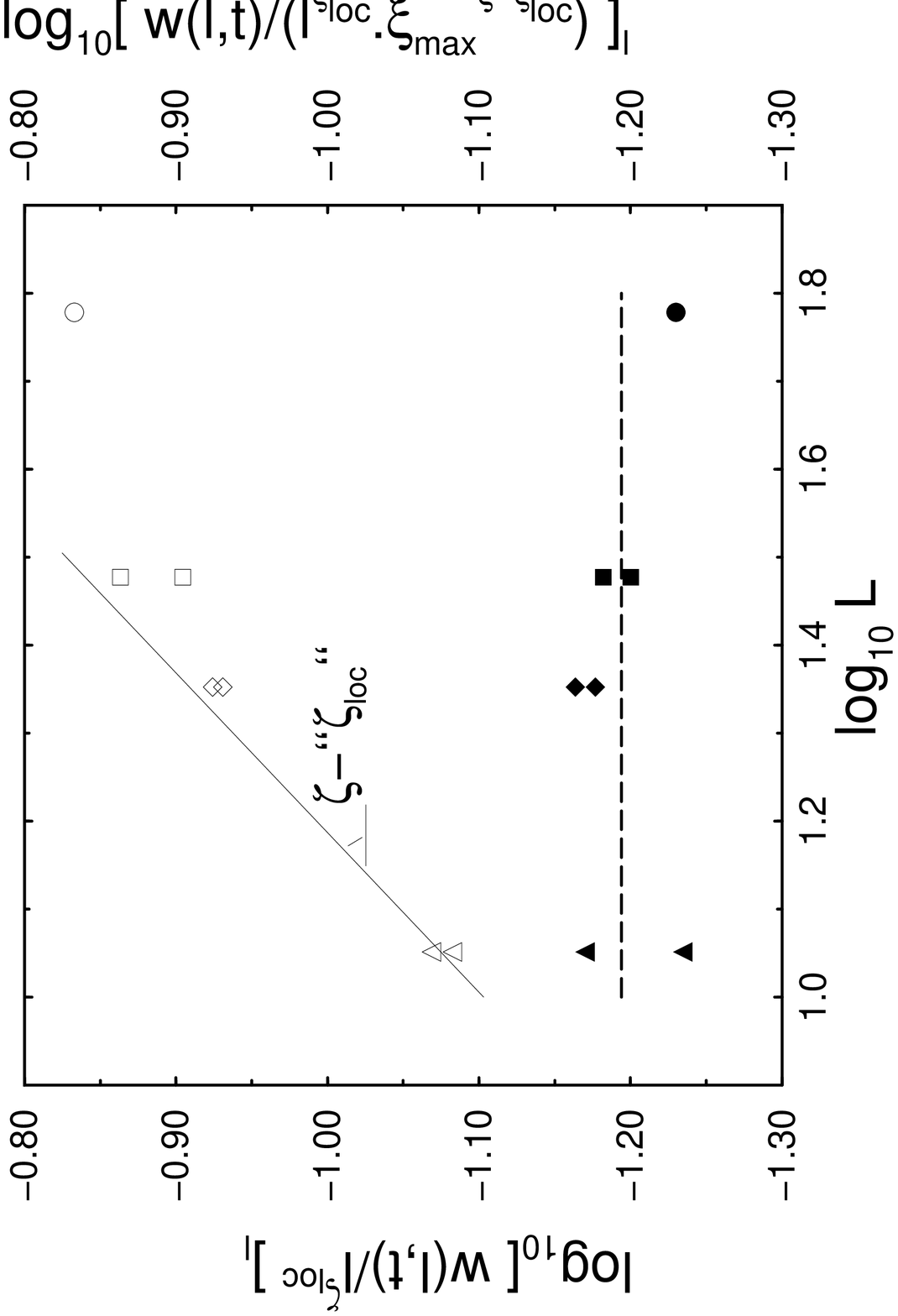}
\end{document}